\documentclass{article}

\PassOptionsToPackage{numbers, compress}{natbib}
\usepackage[preprint]{neurips_2023}

\usepackage[utf8]{inputenc} % allow utf-8 input
\usepackage[T1]{fontenc}    % use 8-bit T1 fonts
\usepackage{hyperref}       % hyperlinks
\usepackage{url}            % simple URL typesetting
\usepackage{booktabs}       % professional-quality tables
\usepackage{amsfonts}       % blackboard math symbols
\usepackage{nicefrac}       % compact symbols for 1/2, etc.
\usepackage{microtype}      % microtypography
\usepackage{xcolor}         % colors

\usepackage{amsmath,amssymb,amsfonts}
\usepackage{algorithmic}
\usepackage{graphicx}
\usepackage{textcomp}
\usepackage{graphicx}
\usepackage{caption}
\usepackage{subcaption}
\usepackage{bbm}
\usepackage{bm}
\usepackage{color}
\usepackage{hyperref}
\usepackage{multirow}
\usepackage{enumitem}
\usepackage{algorithm} 
\usepackage{algorithmic}
\usepackage{enumerate}
\usepackage{amsmath}
\usepackage{booktabs}
\usepackage{multirow}
\usepackage{stmaryrd}
\usepackage{color}
\usepackage{balance}
\usepackage{adjustbox}
\usepackage{mathtools}
\usepackage[normalem]{ulem} 

\usepackage{bm}

\newcommand{\eg}{\textit{e.g.}}

\newcommand{\ie}{\emph{i.e.}}

\title{Large Language Models Make Sample-Efficient Recommender Systems}

\author{
    Jianghao Lin \\
    Shanghai Jiao Tong University \\
    Shanghai, China \\
    chiangel@sjtu.edu.cn \\
    \And 
    Xinyi Dai \\
    Huawei Noah’s Ark Lab \\
    Shenzhen, China \\
    daixinyi5@huawei.com \\
    \And 
    Rong Shan \\
    Shanghai Jiao Tong University \\
    Shanghai, China \\
    shanrong@sjtu.edu.cn \\
    \And 
    Bo Chen \\
    Huawei Noah’s Ark Lab \\
    Shenzhen, China \\
    chenbo116@huawei.com \\
    \And 
    Ruiming Tang \\
    Huawei Noah’s Ark Lab \\
    Shenzhen, China \\
    tangruiming@huawei.com \\
    \And 
    Yong Yu \\
    Shanghai Jiao Tong University \\
    Shanghai, China \\
    yyu@sjtu.edu.cn \\
    \And 
    Weinan Zhang \thanks{Corresponding author.} \\
    Shanghai Jiao Tong University \\
    Shenzhen, China \\
    wnzhang@sjtu.edu.cn \\
}

\begin{document}

\maketitle

\begin{abstract}
Large language models (LLMs) have achieved remarkable progress in the field of natural language processing (NLP), demonstrating remarkable abilities in producing text that resembles human language for various tasks. This opens up new opportunities for employing them in recommender systems (RSs). In this paper, we specifically examine the sample efficiency of LLM-enhanced recommender systems, which pertains to the model's capacity to attain superior performance with a limited quantity of training data. Conventional recommendation models (CRMs) often need a large amount of training data because of the sparsity of features and interactions. Hence, we propose and verify our core viewpoint: Large Language Models Make Sample-Efficient Recommender Systems. We propose a simple yet effective framework (i.e., Laser) to validate the viewpoint from two aspects: (1) LLMs themselves are sample-efficient recommenders; and (2) LLMs, as feature generators and encoders, make CRMs more sample-efficient. Extensive experiments on two public datasets show that Laser requires only a small fraction of training samples to match or even surpass CRMs that are trained on the entire training set, demonstrating superior sample efficiency.
\end{abstract}

% \vspace{-20pt}
\section{Introduction}
\label{sec:intro}

Large language models (LLMs) have achieved remarkable progress in the field of natural language processing (NLP), showing impressive abilities to generate human-like texts for a broad range of tasks~\cite{zhao2023survey,liu2023chatgpt,sun2023chatgpt,dai2023uncovering}. 
Consequently, recent works start to investigate the application of LLMs in recommender systems. 
They adopt LLMs for various recommendation tasks, and show promising performance from different aspects (\eg, user profiling)~\cite{hou2022towards,yu2021tiny,wang2022transrec}. 
In this paper, we mainly focus on promoting the \textbf{sample efficiency} of recommender systems by involving large language models~\cite{lin2024data}.

Sample efficiency refers to the ability of a model to achieve high performance with a limited amount of training data. 
A sample-efficient model can learn effectively from a smaller number of samples, reducing the resources and time required for training. 
In conventional recommender systems, we usually convert raw data into an ID data through one-hot encoding. 
Due to the data sparsity, conventional recommendation models (CRMs) typically require a large volume of training records to achieve satisfactory performance, resulting in sample inefficiency issue~\cite{pan2022exploiting}.

To this end, in this paper, we propose our core viewpoint, \ie, \textbf{\underline{La}rge Language Models Make \underline{S}ample-\underline{E}fficient \underline{R}ecommender Systems} (\textbf{Laser}), where a novel framework named \textbf{Laser} is designed to validate such a core viewpoint from the following two key aspects:
\begin{itemize}[leftmargin=12pt]
    \item \textbf{LLMs themselves are sample-efficient recommenders.}
    \item \textbf{LLMs make conventional recommender systems more sample-efficient.}
\end{itemize}
To the best our knowledge, we are the first to systematically investigate the impact of LLMs on the sample efficiency of recommender systems. 
Experiments show that our proposed Laser framework require only a small fraction of training samples to match or even surpass (CRMs) that are trained on the entire training set.

\section{Related Works}
This paper is closely related to the following two research domains: (1) LLM-enhanced recommender systems, and (2) sample efficiency.

\subsection{\textbf{LLM-enhanced Recommender Systems}}

The emergence of large language models (LLMs) has flourish in the field of natural language processing (NLP), and LLMs are potentially revolutionizing various downstream applications like symbolic reasoning~\cite{jha2023neuro} code analysis~\cite{rahmani2023improving} and recommender systems~\cite{lin2023can}. 
Generally, there are two major paradigms to leverage LLMs to further enhance the performance of recommender systems~\cite{lin2023can}: (1) adapting LLMs themselves as the recommender~\cite{lin2023rella,bao2023tallrec,wang2024towards,wang2023flip}, which corresponds to our Laser$_{LLM\;only}$, and (2) adapting LLMs for the feature engineering or feature encoding~\cite{xi2023towards,fu2023unified,tian2023ufin}, which corresponds to our Laser$_{LLM+CRM}$.

\textbf{LLMs as the recommenders directly.}
In this paradigm, we do not involve conventional recommendation models (CRMs) for the user preference estimation, and should rely on LLMs themselves to directly infer the user's preference towards a target item given his or her behavior history. In this way, we can benefit from the large capacity, open-world knowledge, and emerging abilities like reasoning of LLMs, which deepens the user intention understanding and thereby improves the final recommendation performance. 
According to existing works, in this case, LLMs are adopted to fulfill either the item scoring task~\cite{liu2022ptab,zhang2021language,li2023pbnr,bao2023tallrec,li2023text,zhang2023prompt,mao2023unitrec}, or item generation task~\cite{hua2023up5,geng2023vip5,hua2023index,zhang2023chatgpt,hou2023large,chen2023palr,petrov2023generative,wang2023zero}. 
Also, various works~\cite{geng2022recommendation,cui2022m6,zhang2023recommendation} attempt to utilize the multi-task capacity of LLMs, and
instruct LLMs to solve the multiple tasks (e.g., both scoring and
generation) through a unified language interface.

\textbf{LLMs for the feature engineering and features encoding.} 
In this paradigm, we enhance the performance of traditional recommendation models, which corresponds to our proposed Laser$_{LLM+CRM}$. 
We still maintain the conventional recommendations (CRMs) as the major backbone of the recommender system, and adapt LLMs to generate auxiliary text data~\cite{liu2023first,borisov2022language,li2023taggpt,mysore2023large,carranza2023privacy,christakopoulou2023large} and encode the corresponding textual features~\cite{hou2023learning,zhang2022twhin,fu2023exploring,yuan2023go,qiu2021u,li2023exploring,he2022ptm4tag,muhamed2021ctr} (\eg, item content understanding, user profile modeling), which would serve as the additional inputs for CRMs.
In this way, we are able to omit the online inference of LLMs and make the LLM processing cachable to meet the strict online inference latency constrains of industrial applications.

\subsection{\textbf{Sample Efficiency}}

In the field of machine learning, the sample efficiency of a model refers to its ability to achieve satisfactory performance with a limited amount of training data~\cite{al2015efficient}. 
Improving sample efficiency is crucial for addressing practical challenges in various domains, such as reducing data collection costs~\cite{feng2024sample}, mitigating data annotation efforts~\cite{ding2024data}, and enabling rapid learning in real-world settings where labeled data is scarce~\cite{yu2018towards}. 
In the realm of recommender systems, the sample efficiency is of great importance due to the data sparsity problem~\cite{lin2024data,pan2022exploiting}.
Various approaches have been proposed to improve the sample efficiency of recommender systems, ranging from leveraging auxiliary information through transfer learning and domain adaptation~\cite{zhu2021cross} to employing meta-learning techniques to enable quick adaption to new users or items with limited data~\cite{wang2022deep}. 
Recently, the emerging of large language models have present a promising solution to potentially enhance the sample efficiency of recommender systems with their vast amount of open-world knowledge and powerful reasoning abilities. 
To the best of our knowledge, we are the first to systematically investigate the impact of LLMs on the sample efficiency of recommender systems under two different paradigms, \ie, LLMs as recommenders, and LLMs for feature engineering and encoding.
% \section{Preliminary}

% \vspace{-20pt}
\section{Methodology}

\subsection{\textbf{Preliminary and Overview}}

We focus on a core task of recommender systems, \ie, click-through rate (CTR) prediction, which is a binary classification problem to estimate the click probability $\hat{y}\in[0,1]$ of a user on a target item given a certain context. 
The dataset can be formulated as $\{(x_i,y_i)\}_{i=1}^N$, where $x_i$ is the input and $y_i\in\{1,0\}$ is the label (\ie, click or not)~\cite{lin2023map,lin2024clickprompt}. 

There are two different types of data processing towards the same raw input $x_i$. For one thing, we can perform one-hot encoding to convert $x_i$ into ID input $x_i^{ID}$ for conventional recommendation models. For another, we can adopt hard prompt template to transform $x_i$ into textual input $x_i^{text}$ for LLMs. Accordingly, the label $y_i$ is also converted into the binary answer words $y_i^{text}\in\{``\text{Yes}", ``\text{No}"\}$.

% \subsection{\textbf{Overview of Laser Framework}}

\begin{figure}[t]
    % \vspace{-7pt}
    \centering
    \includegraphics[width=0.7
\textwidth]{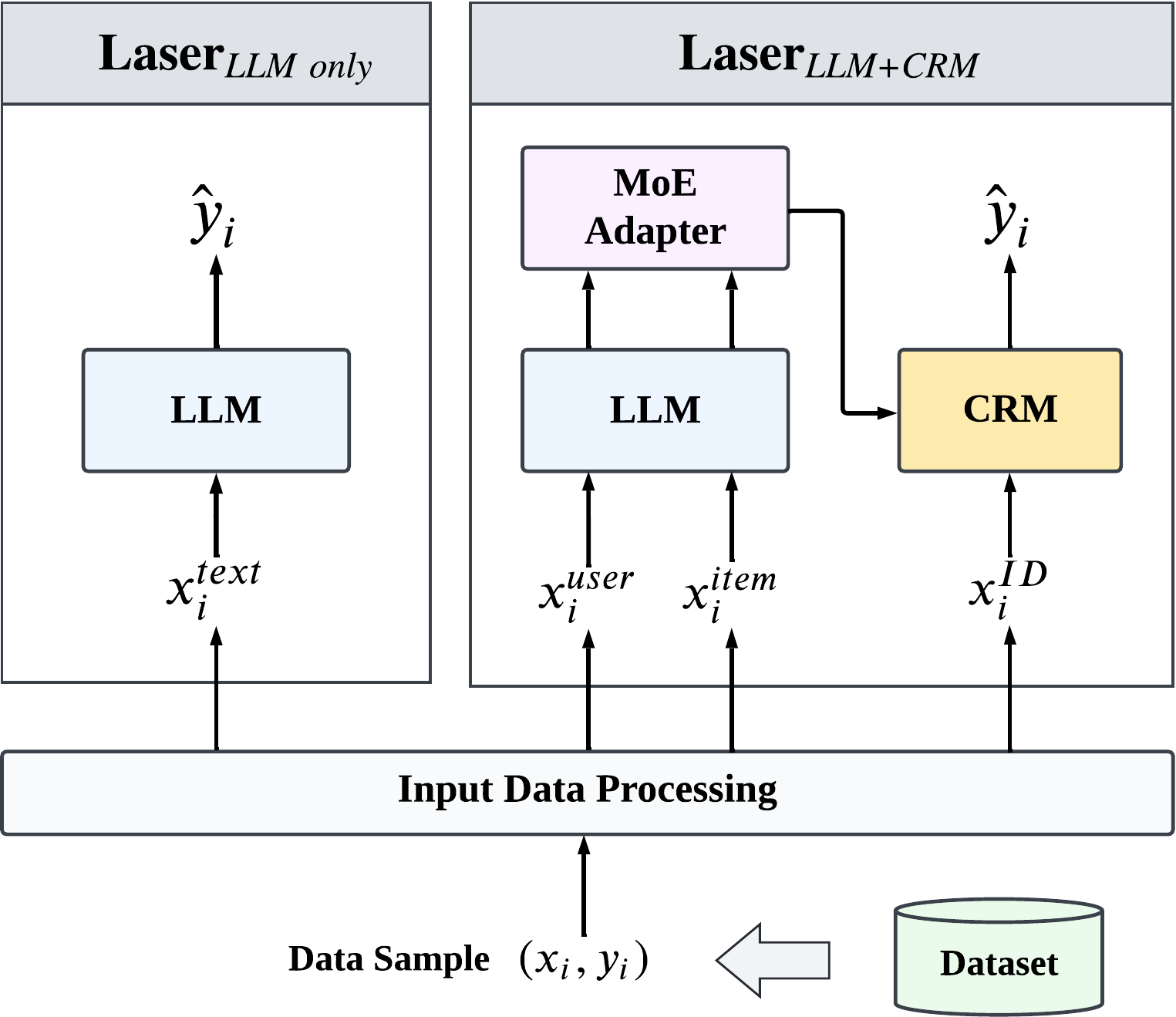}
    \caption{The overall framework of Laser.}
    % \vspace{-5pt}
    \label{fig:framework}
    % \vspace{-5pt}
\end{figure}

We illustrate the overall framework of our proposed Laser in Figure~\ref{fig:framework}. 
The Laser framework consists of two different paradigms to enhance the recommender systems with large language models in terms of sample efficiency:
\begin{itemize}[leftmargin=12pt]
    \item Laser$_{LLM\;only}$ directly adopts a pure large language model as the recommender itself for CTR estimation without involving conventional recommendation models (CRMs). In this paradigm, we require LLMs to answer the binary question about the user's preference towards a target item with specified key words, \ie, Yes or No.
    \item Laser$_{LLM+CRM}$ uses LLMs as the feature generator and encoder to assist CRMs for CTR prediction. In this paradigm, we directly utilize the predicted probability from a conventional recommendation model augmented with textual features from LLMs as inputs.
\end{itemize}

Next, we will elaborate on these two paradigms of LLM-enhanced recommender systems.

\subsection{Laser$_{\textit{LLM\;only}}$: LLM as Recommender Itself}

 % formulates CTR prediction as a binary question answering task by directly adopting LLM as the recommender.
% The paradigm of Laser$_{LLM\;only}$ corresponds to our key statement
When directly adopting LLMs as the recommenders, we formulate CTR prediction as a binary question answering task.
An example of input template is given as follows:
\begin{equation*}
  \tag{A}
  \label{quote:prompt template}
  \parbox{\dimexpr\linewidth-2.5em}{%
    \strut
    $\bm{x^{text}}=\;\,\,$The user watched the following movies in order in the past and rated them: ['0. The Terminator (4 stars)', '1. Back to Future (5 stars)', '2. Matrix (4 stars)']. Please deduce whether the user will like the movie Aliens. You should ONLY tell me yes or no.
    \strut
}
\end{equation*}
% \begin{equation*}
%   \tag{A}
%   \label{quote:prompt template}
%   \parbox{\dimexpr\linewidth-6em}{%
%     \strut
%     $\bm{x^{text}}=\;\,\,$The user watched the following movies in order in the past and rated them: \{\{user\_history\}\}. Please deduce whether the user will like the movie \{\{target movie\}\}. You should ONLY tell me yes or no.
%     \strut
% }
% \end{equation*}
Then, we apply instruction tuning over the LLM with the causal language modeling objective:
\begin{equation}
    \max_{\Theta}\sum\nolimits_{(x^{text},y^{text})\in\mathcal{D}}\, \sum\nolimits_{j=1}^{|y^{text}|}\log P_{\Theta}(y^{text}_{j}|x^{text},y_{<j}^{text}),
\end{equation}
where $\Theta$ is the parameters of the LLM, $\mathcal{D}$ is the training set, $y^{text}_j$ is the $j$-th token of the textual output $y^{text}$, and $y^{text}_{<j}$ denotes the tokens before $y^{text}_j$. 
During evaluation, since we want a continuous click probability in $[0,1]$ rather than a generated discrete token (\ie, Yes or No), we perform a bidimensional softmax over the logits of ``Yes'' and ``No'' tokens (denoted as $s_{yes}$ and $s_{no}$) to obtain the estimated click probability $\hat{y}$:
\begin{equation}
    \hat{y}=\frac{\exp(s_{yes})}{\exp(s_{yes})+\exp(s_{no})}\in (0,1).
\label{eq:llm_pred}
\end{equation}

In this paradigm, we do not involve the conventional recommendation models like DIN~\cite{zhou2018deep} and SIM~\cite{pi2020search}, and only adopt LLM itself for both instruction tuning and evaluation. 
It is worth noting that such an estimated click-through rate $\hat{y}_i$ in Eq.~\ref{eq:llm_pred} is only leveraged for the evaluation on the testing set. 
We preserve the common instruction tuning and causal language modeling paradigm for LLMs during the training phase.

\subsection{Laser$_{\textit{LLM+CRM}}$: LLM as Feature Enhancer for CRM}

Laser$_{LLM+CRM}$ uses LLMs as the feature generators and encoders to assist CRMs in a model-agnostic manner.
Specifically, for input $x_i$, we construct textual instructions ($x_i^{user}$ and $x_i^{item}$) to let LLM infer and generate the knowledge for both user-side preference and item-side facts.
After generation, we apply mean pooling over the hidden states to get the knowledge vectors for both user and item: 
% \cb{Bold for the vectors?  e.g., $\mathbf{h}^{user}_i$, $\boldsymbol{\alpha}_i^{*}$, $\mathbf{z}_i^{user}$}
\begin{equation}
    h^{user}_i=\operatorname{LLM}(x_i^{user}),\;\;h^{item}_i=\operatorname{LLM}(x_i^{item}).
\end{equation}

To ensure the dimensional consistency and bridge the latent space between language and recommendation, we design two parallel mixture-of-experts (MoE) adapters for user and item knowledge vectors, respectively:
\begin{equation}
\begin{aligned}
    \alpha_i^{*}&=\operatorname{Softmax}(g^{*}(h_i^{*})),\;*\in\{user, item\},\\
    z_i^{*}&=\sum\nolimits_{j=1}^{J}\alpha^{*}_{i,j}\times e_j^{*}(h_i^{*}),\;*\in\{user, item\},
\end{aligned}
\end{equation}
where $J$ is the number of expert network. The gating networks $g^{*}(\cdot)$ and expert networks $e^{*}(\cdot)$ are all designed as multi-layer perceptrons (MLPs).
Next, we can feed the LLM-augmented user \& item representations $z_i^{user}$ and $z_i^{item}$ as additional input features into arbitrary CRMs to perform CTR estimation training and inference:
\begin{equation}
    \hat{y}_i=\operatorname{CRM}(x_i^{ID},z_i^{user},z_i^{item}).
\end{equation}

In this paradigm, we utilize the predicted probability from a conventional recommendation model with augmented textual features from LLMs as inputs. In this way, the LLM processing for the augmented features is thereby cachable and can be conducted offline to satisfy the strict online inference latency for recommender systems.

\subsection{\textbf{Discussion}}

We would like to provide further discussion from the following two perspectives about our proposed Laser framework: sample efficiency and inference latency.

\vspace{10pt}
\noindent\textbf{Sample Efficiency.}
LLMs possess extensive open-world knowledge and powerful logical reasoning capabilities. 
This enables them to infer the user preferences and factual knowledge about items under few-shot or even zero-shot settings. 
The formation of such prior knowledge within the recommender system (both on the user and item sides) can help the model to better understand and leverage the limited interaction data, and thus help alleviate sample inefficiency issue caused by feature and interaction sparsity, reducing our reliance on the quantity of training samples.
Therefore, our proposed Laser gains better sample efficiency compared with CRMs. 
Note that we only discuss the sample efficiency for the recommendation training stage. The pretraining stage of LLMs is no longer considered since LLMs are actually off-the-shelf for various downstream applications.

\vspace{10pt}
\noindent\textbf{Inference Latency.}
The inference latency of Laser$_{LLM\;only}$ is fairly high and unavoidable for practical recommender systems, since it directly uses LLMs to perform per-sample CTR estimation task. 
However, for Laser$_{LLM+CRM}$, we decompose the per-sample inference into user-level and item-level knowledge generation and encoding.
The LLM-augmented representations $z_i^{user}$ and $z_i^{item}$ from LLM can therefore be prestored to avoid online real-time calls to LLM.
Hence, the inference time of Laser$_{LLM+CRM}$ is nearly the same as the original CRM backbone, which satisfies the strict inference latency constraint for real-world practical recommender systems. 
Specifically, assume the inference time complexity of the backbone CRM is $O(F(n, m))$, where $n$ is the number of feature fields and $m$ is the embedding size, and $F(\cdot,\cdot)$ is a polynomial function. Then, the time complexity of Laser$_{LLM+CRM}$ is $O(F(n+l,m))=O(f(n,m))$ if $l<<n$, which is equivalent to the complexity of the backbone CRM. 

\section{Experiment}

\subsection{\textbf{Experiment Setups}}

We conduct experiments on two real-world public datasets, \ie, \href{https://grouplens.org/datasets/book-crossing/}{BookCrossing} and \href{https://grouplens.org/datasets/movielens/1m/}{MovieLens-1M}. 
We adopt three types of baseline models: (1) feature interaction based CTR models, including DeepFM, AutoInt, DCNv2; (2) sequential CTR models, including GRU4Rec, Caser, SASRec, DIN, SIM; (3) language model enhanced CTR models, including CTR-BERT, PTab, P5.
We use AUC and LogLoss as the evaluation metrics. 
For Laser, as default, we select Vicuna-13B~\cite{vicuna2023} as the LLM backbone and SIM as the CRM backbone. 
SIM is the best baseline CRM, and we would choose different LLMs to analyze the compatibility of Laser.
For fair comparison with SIM, we collect the user history behaviors with the same retrieval techniques.
While all the baselines are trained on the entire training set, we randomly sample 10\% and 50\% data of the training set for Laser$_{LLM\;only}$ and Laser$_{LLM+CRM}$, respectively. 
% All the experiments are exclusively conducted on the same server with NVIDIA Tesla V100 GPUs.
% \xy{can we show the performance of CRM trained with same proportion of training data? To demonstrate sample efficiency, it is better to show experimental results with three or more different data sizes.}

\begin{table}[t]
\centering
% \vspace{-10pt}
\caption{Overall performance. 
The best result is in \textbf{bold}. 
The second-best value is \underline{underlined}. \uwave{Wavy underline} denotes the third-best value.
Rel.Impr indicates the relative AUC improvement of Laser$_{LLM\;only}$ against baselines. 
* indicates statistically
significant improvement with $p$ < 0.001.
}
% \vspace{-10pt}
\label{tab:performance}
\resizebox{0.99\textwidth}{!}{
\renewcommand\arraystretch{1.1}
\begin{tabular}{c|ccc|ccc}
\toprule
\hline
\multicolumn{1}{c|}{\multirow{2}{*}{Model}} & \multicolumn{3}{c|}{BookCrossing} & \multicolumn{3}{c}{MovieLens-1M} \\ 
\multicolumn{1}{c|}{} & AUC $\uparrow$  & Log Loss $\downarrow$ & Rel.Impr & AUC $\uparrow$  & Log Loss $\downarrow$ & Rel.Impr\\ 
   \hline 
DeepFM & 0.7496 & 0.5953 & 1.05\% & 0.7915 & 0.5484 & 1.49\% \\ 
AutoInt & 0.7481 & 0.6840 & 1.26\% & 0.7929 & 0.5453 & 1.31\% \\ 
DCNv2 & 0.7472 & 0.6816 & 1.38\% & 0.7931 & 0.5464 & 1.29\%\\ 
GRU4Rec & 0.7479 & 0.5930 & 1.28\% & 0.7926 & 0.5453 & 1.35\% \\ 
Caser & 0.7478 & 0.5990 & 1.30\% & 0.7918 & 0.5464 & 1.45\% \\ 
SASRec & 0.7482 & 0.5934 & 1.24\% & 0.7934 & 0.5460 & 1.25\% \\ 
DIN & 0.7477 & 0.6811 & 1.31\% & 0.7962 & 0.5425 & 0.89\% \\ 
SIM & \underline{0.7541} & \underline{0.5893} & 0.45\% & \uwave{0.7992} & \uwave{0.5387} & 0.51\% \\ 
CTR-BERT & 0.7448 & 0.5938 & 1.71\% & 0.7931 & 0.5457 & 1.29\% \\ 
PTab & 0.7429 & 0.6154 & 1.97\% & 0.7955 & 0.5428 & 0.98\% \\ 
P5 & 0.7438 & 0.6128 & 1.84\% & 0.7937 & 0.5478 & 1.21\% \\ 
\hline
Laser$_{LLM\;only}$ & \textbf{0.7575}* & \uwave{0.5919} & - & \textbf{0.8033}* & \textbf{0.5362}* & - \\ 
Laser$_{LLM+CRM}$ & \uwave{0.7508} & \textbf{0.5848}* & - & \underline{0.7996} & \underline{0.5375} & - \\ 
  
   \hline  
   \bottomrule          
\end{tabular}
}
% \vspace{-5pt}
\end{table}

\begin{table}[t]
\caption{The inference time (second) per sample.
}
% \vspace{-5pt}
\centering
\label{tab:inference}
\resizebox{0.95\textwidth}{!}{
\renewcommand\arraystretch{1.1}
\begin{tabular}{c|ccccc}
\toprule
\hline
Dataset & DCNv2 & SIM & PTab & Laser$_{LLM\;only}$ & Laser$_{LLM+CRM}$ \\
\hline
BookCrossing & $2.34\times 10^{-4}$ & $2.45\times 10^{-4}$ & $5.23\times 10^{-3}$ & $7.89\times 10^{-1}$ & $2.77\times 10^{-4}$ \\
MovieLens-1M & $1.65\times10^{-4}$ & $2.07\times10^{-4}$ & $3.95\times10^{-3}$ & $7.42\times 10^{-1}$ & $2.34\times 10^{-4}$ \\
   \hline  
   \bottomrule          
\end{tabular}
}
% \vspace{-10pt}
\end{table}

\subsection{\textbf{Experiment Results}}

We report the performance of Laser and baselines in Table~\ref{tab:performance}, from which we can draw the following observations: 
\begin{itemize}[leftmargin=12pt]
    \item With only a small fraction of training samples, both Laser$_{LLM\;only}$ (10\% data) and Laser$_{LLM+CRM}$ (50\% data) are able to match or even surpass CRMs that are trained on the entire training set. This demonstrates that LLMs are not only inherently sample-efficient recommenders, but can also enhance the sample efficiency of CRMs.
    \item The sample efficiency of Laser$_{LLM+CRM}$ is inferior to that of Laser$_{LLM\;only}$ due to the involvement of CRMs, which increases the demand on the quantity of training samples. 
    \item To make the generated knowledge cachable to satisfy the inference latency, Laser$_{LLM+CRM}$ merely leverages LLMs to generate open-world knowledge at user and item levels, which lacks mutual interactions between the users and items. On the contrary, Laser$_{LLM\;only}$ conducts the user preference modeling at sample level (\ie, user-item pairs), which provides finer-grained interactions and results in better recommendation performance.
    \item Language model enhanced CTR baselines perform poorly since they only incorporate small language models like BERT. The superior performance and sample efficiency is closely related to the model size and emergent abilities (\eg, reasoning) of the involved language models.
\end{itemize}

Then, we report the inference time per sample with batch size as 128 for Laser, as well as baselines, to investigate whether they are practical for real-world applications. 
We choose DCNv2, SIM and PTab as representative models for the three types of baselines, and give the results in Table~\ref{tab:inference}, from which we obtain the following observations:
\begin{itemize}[leftmargin=12pt]
    \item Despite the superior performance and sample efficiency Laser$_{LLM\;only}$ achieves, its inference overhead generally violates the latency constraint of real-world recommender systems, where each user request should be responded within tens of milliseconds.
    \item By prestoring the LLM-augmented user and item representations, Laser$_{LLM+CRM}$ is able to achieve nearly the same inference latency as other CRMs, which definitely satisfies the inference latency constraint. This shows that, with the help of LLMs and proper model design, we can achieve sample-efficient recommender systems for real-world applications.
\end{itemize}

\begin{table}[h]
    % \vspace{-5pt}
    \caption{The compatibility of Laser w.r.t. different backbone LLMs on MovieLens-1M dataset.
    }
    % \vspace{-5pt}
    \centering
    \label{tab:generalization}
    \resizebox{0.95\textwidth}{!}{
    \renewcommand\arraystretch{1.1}
    \begin{tabular}{c|c|cc|cc|cc}
    \toprule
    \hline
    \multicolumn{1}{c|}{\multirow{2}{*}{Metric}} & \multicolumn{1}{c|}{\multirow{2}{*}{SIM}} & \multicolumn{2}{c|}{\multirow{1}{*}{Mistral-7B}} & \multicolumn{2}{c|}{\multirow{1}{*}{Vicuna-7B}} & \multicolumn{2}{c}{\multirow{1}{*}{Vicuna-13B}} \\ 
    \multicolumn{1}{c|}{} & \multicolumn{1}{c|}{} & LLM only & LLM+CRM & LLM only & LLM+CRM & LLM only & LLM+CRM \\
   \hline 
    AUC & 0.7992 & 0.8005 & 0.7990 & 0.8016 & 0.7997 & 0.8033 & 0.7996 \\
    LogLoss & 0.5387 & 0.5388 & 0.5385 & 0.5365 & 0.5372 & 0.5362 & 0.5375 \\
   \hline  
   \bottomrule          
\end{tabular}
}
\end{table}

\noindent Next, we investigate the compatibility of Laser w.r.t. LLM backbones with different model architectures and sizes on MovieLens-1M dataset. We choose Mistral-7B, Vicuna-7B, and Vicuna-13B as the LLM backbones. The results are given in Table~\ref{tab:generalization}. 
We can observe that the sample efficient property commonly exists for various different LLMs under our proposed Laser framework, where Laser can generally achieve better performance compared with SIM to a sampled small fraction of training samples.

\section{Conclusion}

This paper investigates the sample efficiency property of recommender systems enhanced by large language models. 
We propose a simple yet effective framework (\ie, Laser) to validate the core viewpoint - \textit{large language models make sample-efficient recommender systems} - from two aspects: (1) LLMs themselves are sample-efficient recommenders; and (2) LLMs make conventional recommender systems more sample-efficient.
Experiments show that, with only a small fraction of training samples, our proposed Laser can match or even surpass conventional recommendation models that are trained on the entire training set.
For future work, we aim to improve the sample efficiency of LLM-based recommender systems from the following two aspects: (1) exploring effective strategy to select the few-shot training samples instead of uniformly sampling, and (2) applying Laser for downstream applications like code snippet recommendation.

\bibliographystyle{plainnat}
\bibliography{fcs}

\end{document}